\documentclass[12pt] {article}
\begin{document}
\setcounter{page}{1}
\def\theequation{\arabic{section}.\arabic{equation}}
\def\theequation{\thesection.\arabic{equation}}
\setcounter{section}{0}

\title{On the $\Delta\Delta$ component of the deuteron in
the Nambu--Jona--Lasinio model of light nuclei}

\author{A. N. Ivanov~\thanks{E--mail: ivanov@kph.tuwien.ac.at, Tel.:
+43--1--58801--14261, Fax: +43--1--58801--14299}~${\textstyle
^\ddagger}$, H. Oberhummer~\thanks{E--mail: ohu@kph.tuwien.ac.at,
Tel.: +43--1--58801--14251, Fax: +43--1--58801--14299} ,
N. I. Troitskaya~\thanks{Permanent Address: State Technical
University, Department of Nuclear Physics, 195251 St. Petersburg,
Russian Federation} , M. Faber~\thanks{E--mail:
faber@kph.tuwien.ac.at, Tel.: +43--1--58801--14261, Fax:
+43--1--58801--14299}}

\date{\today}

\maketitle

\begin{center}
{\it Institut f\"ur Kernphysik, Technische Universit\"at Wien,\\
Wiedner Hauptstr. 8-10, A-1040 Vienna, Austria}
\end{center}

\begin{center}
\begin{abstract}
The probability $P(\Delta\Delta)$ to find the $\Delta\Delta$ component
inside the deuteron, where $\Delta$ stands for the $\Delta(1232)$
resonance, is calculated in the Nambu--Jona--Lasinio model of light
nuclei.  We obtain $P(\Delta\Delta) = 0.3\,\%$. This prediction
agrees good with the experimental estimate $P(\Delta\Delta) < 0.4\,\%$
at 90$\%$ of CL (D. Allasia {\it et al.}, Phys. Lett. B174 (1986)
450).
\end{abstract}
\end{center}

\begin{center}

\end{center}

\newpage

\section{Introduction}
\setcounter{equation}{0}

\hspace{0.2in} As has been stated in Ref.[1] that nowadays there is a
consensus concerning the existence of non--nucleonic degrees of
freedom in nuclei. The non--nucleonic degrees of freedom can be
described either within QCD in terms of quarks and gluons [2] or in
terms of mesons and nucleon resonances [3].

In this letter we investigate the non--nucleonic degrees of freedom in
terms of the $\Delta(1232)$ resonance and calculate the contribution
of the $\Delta\Delta$ component to the deuteron in the
Nambu--Jona--Lasinio model of light nuclei or differently the nuclear
Nambu--Jona--Lasinio (NNJL) model [4,5]. As has been shown in
Ref.\,[4] the NNJL model is motivated by QCD.  The deuteron appears
{\it in the nuclear phase of QCD} as a neutron--proton collective
excitation, the Cooper np--pair, induced by a phenomenological local
four--nucleon interaction. The NNJL model describes low--energy
nuclear forces in terms of one--nucleon loop exchanges providing a
minimal transfer of nucleon flavours from initial to final nuclear
states and accounting for contributions of nucleon--loop anomalies
which are completely determined by one--nucleon loop diagrams. The
dominance of contributions of nucleon--loop anomalies to effective
Lagrangians of low--energy nuclear interactions is justified in the
large $N_C$ expansion, where $N_C$ is the number of quark colours
[4]. As has been shown in Refs.\,[5] the NNJL model describes good
low--energy nuclear forces for electromagnetic and weak nuclear
reactions with the deuteron of astrophysical interest such as the
neutron--proton radiative capture n + p $\to$ D + $\gamma$, the
solar proton burning p + p $\to$ D + e$^+$ + $\nu_{\rm e}$, the
pep--process p + e$^-$ + p $\to$ D + $\nu_{\rm e}$ and reactions of
the disintegration of the deuteron by neutrinos and anti--neutrinos
caused by charged $\nu_{\rm e}$ + D $\to$ e$^-$ + p + p,
$\bar{\nu}_{\rm e}$ + D $\to$ e$^+$ + n + n and neutral $\nu_{\rm
e}(\bar{\nu}_{\rm e})$ + D $\to$ $\nu_{\rm e}(\bar{\nu}_{\rm e})$ + n
+ p weak currents.

A phenomenological Lagrangian of the ${\rm n p D}$ interaction is
defined by [4]
\begin{eqnarray}\label{label1.1}
{\cal L}_{\rm npD}(x) = - ig_{\rm V}[\bar{p}(x)\gamma^{\mu}n^c(x) -
\bar{n}(x)\gamma^{\mu}p^c(x)] D_{\mu}(x),
\end{eqnarray}
where $D_{\mu}(x)$, $n(x)$ and $p(x)$ are the interpolating fields of
the deuteron, the neutron and the proton. The phenomenological
coupling constant $g_{\rm V}$ is related to the electric quadrupole
moment of the deuteron $Q_{\rm D} = 0.286\,{\rm fm}$: $g^2_{\rm V} =
2\pi^2 Q_{\rm D}M^2_{\rm N}$ [4], where $M_{\rm N} = 940\,{\rm MeV}$
is the nucleon mass. In the isotopically invariant form the
phenomenological interaction  Eq.(\ref{label1.1}) can be written as
\begin{eqnarray}\label{label1.2}
{\cal L}_{\rm npD}(x) = g_{\rm
V}\,\bar{N}(x)\gamma^{\mu}\tau_2N^c(x)\,D_{\mu}(x),
\end{eqnarray}
where $\tau_2$ is the Pauli isotopical matrix and $N(x)$ is a doublet
of a nucleon field with components $N(x) = (p(x), n(x))$, $N^c(x) =
C\,\bar{N}^T(x)$ and $\bar{N^c}(x) = N^T(x)\,C$, where $C$ is a charge
conjugation matrix and $T$ is a transposition.

In the NNJL model [5] the $\Delta(1232)$ resonance is the
Rarita--Schwinger field [6] $\Delta^a_{\mu}(x)$, the isotopical index
$a$ runs over $a = 1,2,3$, having the following free Lagrangian
[7,8]: 
\begin{eqnarray}\label{label1.3}
\hspace{-0.5in} {\cal L}^{\Delta}_{\rm kin}(x) =
\bar{\Delta}^a_{\mu}(x) [-(i\gamma^{\alpha}\partial_{\alpha} -
M_{\Delta}) \,g^{\mu\nu} +
\frac{1}{4}\gamma^{\mu}\gamma^{\beta}(i\gamma^{\alpha}\partial_{\alpha}
- M_{\Delta}) \gamma_{\beta}\gamma^{\nu}] \Delta^a_{\nu}(x),
\end{eqnarray}
where $M_{\Delta} = 1232\,{\rm MeV}$ is the mass of the $\Delta(1232)$
resonance field $\Delta^a_{\mu}(x)$. In terms of the eigenstates of
the electric charge operator the fields $\Delta^a_{\mu}(x)$ are given
by [7,8]
\begin{eqnarray}\label{label1.4}
\begin{array}{llcl}
&&\Delta^1_{\mu}(x) = \frac{1}{\sqrt{2}}\Biggr(\begin{array}{c}
\Delta^{++}_{\mu}(x) - \Delta^0_{\mu}(x)/\sqrt{3} \\
\Delta^+_{\mu}(x)/\sqrt{3} - \Delta^-_{\mu}(x)
\end{array}\Biggl)\,,\,
\Delta^2_{\mu}(x) = \frac{i}{\sqrt{2}}\Biggr(\begin{array}{c}
\Delta^{++}_{\mu}(x) + \Delta^0_{\mu}(x)/\sqrt{3} \\
\Delta^+_{\mu}(x)/\sqrt{3} + \Delta^-_{\mu}(x)
\end{array}\Biggl)\,,\\
&&\Delta^3_{\mu}(x) = -\sqrt{\frac{2}{3}}\Biggr(\begin{array}{c}
\Delta^+_{\mu}(x) \\ \Delta^0_{\mu}(x) \end{array}\Biggl).
\end{array}
\end{eqnarray}
The fields $\Delta^a_{\mu}(x)$ obey the subsidiary constraints:
$\partial^{\mu}\Delta^a_{\mu}(x) = \gamma^{\mu}\Delta^a_{\mu}(x) = 0$
[7--9]. The Green function of the free $\Delta$--field is determined by
\begin{eqnarray}\label{label1.5}
\hspace{-0.5in}<0|{\rm T}(\Delta_{\mu}(x_1)\bar{\Delta}_{\nu}(x_2))|0>
= - i S_{\mu\nu}(x_1 - x_2).
\end{eqnarray}
In the momentum representation $S_{\mu\nu}(x)$ reads [5--8]:
\begin{eqnarray}\label{label1.6}
\hspace{-0.5in}S_{\mu\nu}(p) = \frac{1}{M_{\Delta} - \hat{p}}\Bigg( -
g_{\mu\nu} + \frac{1}{3}\gamma_{\mu}\gamma_{\nu} +
\frac{1}{3}\frac{\gamma_{\mu}p_{\nu} -
\gamma_{\nu}p_{\mu}}{M_{\Delta}} +
\frac{2}{3}\frac{p_{\mu}p_{\nu}}{M^2_{\Delta}}\Bigg).
\end{eqnarray}
The most general form of the ${\rm \pi N \Delta}$ interaction
compatible with the requirements of chiral symmetry reads [7]:
\begin{eqnarray}\label{label1.7}
\hspace{-0.5in}&&{\cal L}_{\rm \pi N \Delta}(x) = \frac{g_{\rm \pi
N\Delta}}{2M_{\rm
N}}\bar{\Delta}^a_{\omega}(x)\Theta^{\omega\varphi}N(x)
\partial_{\varphi}\pi^a(x)
+ {\rm h.c.} = \nonumber\\
\hspace{-0.5in}&&= \frac{g_{\rm \pi N\Delta}}{\sqrt{6}M_{\rm
N}}\Bigg[\frac{1}{\sqrt{2}}\bar{\Delta}^+_{\omega}(x)\Theta^{\omega\varphi}
n(x) \partial_{\varphi}\pi^+(x) -
\frac{1}{\sqrt{2}}\bar{\Delta}^0_{\omega}(x)\Theta^{\omega\varphi}
p(x) \partial_{\varphi}\pi^-(x)\nonumber\\
\hspace{-0.5in}&&- \bar{\Delta}^+_{\omega}(x)\Theta^{\omega\varphi}
p(x) \partial_{\varphi}\pi^0(x) -
\bar{\Delta}^0_{\omega}(x)\Theta^{\omega\varphi} p(x)
\partial_{\varphi}\pi^0(x) + \ldots \Bigg],
\end{eqnarray}
where $\pi^a(x)$ is the pion field with the components $\pi^1(x) =
(\pi^-(x) + \pi^+(x))/\sqrt{2}$, $\pi^2(x) = (\pi^-(x) -
\pi^+(x))/i\sqrt{2}$ and $\pi^3(x) = \pi^0(x)$. The tensor
$\Theta^{\omega\varphi}$ is given in Ref.\,[7]:
$\Theta^{\omega\varphi} = g^{\omega\varphi} - (Z +
1/2)\gamma^{\omega}\gamma^{\varphi}$, where the parameter $Z$ is
arbitrary. The parameter $Z$ defines the ${\rm \pi N \Delta}$ coupling
off--mass shell of the $\Delta(1232)$ resonance. There is no consensus
on the exact value of $Z$. From theoretical point of view $Z=1/2$ is
preferred [7].  Phenomenological studies give only the bound $|Z| \le
1/2$ [9]. The value of the coupling constant $g_{\rm \pi N\Delta}$
relative to the coupling constant $g_{\rm \pi NN}$ is $g_{\rm \pi
N\Delta} = 2\,g_{\rm \pi NN}$ [10]. As has been shown in Ref.\,[5] for
the description of the experimental value of the cross section for the
neutron--proton radiative capture for thermal neutrons the parameter
$Z$ should be equal to $Z = 0.473$.  This agrees with the experimental
bound [9]. At $Z=1/2$ we get the result agreeing with the experimental
value of the cross section for the neutron--proton radiative capture
with accuracy about 3$\%$ [5].

For the subsequent calculations of the $\Delta\Delta$ component of the
deuteron it is useful to have the Lagrangian of the ${\rm \pi N
\Delta}$ interaction taken in the equivalent form
\begin{eqnarray}\label{label1.8}
{\cal L}_{\rm \pi N \Delta}(x) = \frac{g_{\rm \pi
N\Delta}}{2M_{\rm N}}\,\partial_{\varphi}\pi^a(x)\bar{N^c}(x)
\Theta^{\varphi\omega}\Delta^a_{\omega}(x)^c + {\rm h.c.},
\end{eqnarray}
where $\Delta^a_{\omega}(x)^c = C \bar{\Delta}^a_{\omega}(x)^T$. Now
we can proceed to the evaluation of the $\Delta\Delta$ component of
the deuteron.

\section{Effective ${\rm \Delta\Delta D}$ interaction}
\setcounter{equation}{0}

\hspace{0.2in} In the NNJL model the existence of the $\Delta\Delta$
component of the deuteron we can understand in terms of the coupling
constants of the effective ${\rm \Delta\Delta D}$ interaction.

In order to evaluate the Lagrangian of the effective ${\rm
\Delta\Delta D}$ interaction ${\cal L}^{\rm \Delta\Delta D}_{\rm eff}(x)$ we
have to obtain, first, the effective Lagrangian of the transition N +
N $\to$ $\Delta$ + $\Delta$. This effective Lagrangian we define in
the one--pion exchange approximation [5,11]
\begin{eqnarray}\label{label2.1}
\int d^4x\,{\cal L}^{\rm NN \to \Delta\Delta}_{\rm
eff}(x)&=&-\frac{g^2_{\rm \pi N\Delta}}{8M^2_{\rm N}}\int\!\!\!\int
d^4x_1\,d^4x_2\,[\bar{\Delta}^a_{\alpha}(x_1)
\Theta^{\alpha\beta}N(x_1)]\,\nonumber\\ &&\times
\frac{\partial}{\partial x^{\beta}_1}\frac{\partial}{\partial
x^{\varphi}_1}[\delta^{ab}\Delta(x_1-x_2)]\,[\bar{N^c}(x_2)
\Theta^{\varphi\omega}\Delta^b_{\omega}(x_2)^c],
\end{eqnarray}
where $\Delta(x_1-x_2)$ is the Green function of $\pi$--mesons.  In
terms of the Lagrangians of the ${\rm n p D}$ interaction and the N +
N $\to$ $\Delta$ + $\Delta$ transition the Lagrangian of the effective
${\rm \Delta\Delta D}$ interaction can be defined by
\begin{eqnarray}\label{label2.2}
&&\int d^4x\,{\cal L}^{\rm \Delta\Delta D}_{\rm eff}(x) = -i\,g_{\rm
V}\,\frac{g^2_{\rm \pi N\Delta}}{4M^2_{\rm N}}\int
d^4x\,d^4x_1\,d^4x_2\, D_{\mu}(x)\,\nonumber\\
&&[\bar{\Delta}^a_{\alpha}(x_1)
\Theta^{\alpha\beta}S_F(x-x_1)\gamma^{\mu}\tau_2S^c_F(x-x_2)
\Theta^{\varphi\omega}\Delta^a_{\omega}(x_2)^c]\,\frac{\partial}{\partial
x^{\beta}_1}\frac{\partial}{\partial x^{\varphi}_1}\Delta(x_1-x_2),
\end{eqnarray}
where $S_F(x-x_1)$ and $S^c_F(x-x_2)$ are the Green functions of the
free nucleon and anti--nucleon fields, respectively. 

Such a definition of the contribution of the $\Delta\Delta$ component
to the deuteron is in agreement with that given by Niephaus {\it et
al.} [12] in the potential model approach (PMA).

For the evaluation of the effective Lagrangian ${\cal L}^{\rm
\Delta\Delta D}_{\rm eff}(x)$ we would follow the large $N_C$
expansion approach to non--perturbative QCD [4]. In the large $N_C$
approach to non--perturbative QCD with $SU(N_C)$ gauge group at $N_C
\to \infty$ the nucleon mass is proportional to the number of quark
colour degrees of freedom, $M_{\rm N} \sim N_C$ [13]. It is
well--known that for the evaluation of effective Lagrangians all
momenta of interacting particles should be kept off--mass shell. This
implies that at leading order in the large $N_C$ expansion
corresponding the $1/M_{\rm N}$ expansion of the momentum integral
defining the effective Lagrangian ${\cal L}^{\rm \Delta\Delta D}_{\rm
eff}(x)$ one can neglect the momenta of interacting particles with
respect to the mass of virtual nucleons. As a result the effective
Lagrangian ${\cal L}^{\rm \Delta\Delta D}_{\rm eff}(x)$ reduces itself
to the local form and reads
\begin{eqnarray}\label{label2.3}
{\cal L}^{\rm \Delta\Delta D}_{\rm eff}(x) =
\frac{g_{\rm V}}{16\pi^2}\,\frac{g^2_{\rm \pi
N\Delta}}{4M^2_{\rm N}}
[\bar{\Delta}^a_{\alpha}(x)\,\Theta^{\alpha\mu\omega}\tau_2
\Delta^a_{\omega}(x)^c]\,D_{\mu}(x),
\end{eqnarray}
where the structure function $\Theta^{\alpha\mu\omega}$ is given by
the momentum integral
\begin{eqnarray}\label{label2.4}
\Theta^{\alpha\mu\omega}=\int\frac{d^4k}{\pi^2i}\,\frac{1}{M^2_{\pi} -
k^2}\, \Theta^{\alpha\beta}k_{\beta}\frac{1}{M_{\rm N} -
\hat{k}}\gamma^{\mu}\frac{1}{M_{\rm N} +
\hat{k}}k_{\varphi}\Theta^{\varphi\omega}.
\end{eqnarray}
Integrating over $k$ we obtain
\begin{eqnarray}\label{label2.5}
\Theta^{\alpha\mu\omega} &=& \frac{1}{3}\,\Big[I_1(M_{\rm N}) -
\frac{5}{2}\,M^2_{\rm N}\,I_2(M_{\rm
N})\Big]\,\Theta^{\alpha\beta}\gamma^{\mu}{\Theta_{\beta}}^{\omega}\nonumber\\
&-&\frac{1}{12}\Big[I_1(M_{\rm N}) - M^2_{\rm N}\,I_2(M_{\rm
N})\Big]\,(\Theta^{\alpha\beta}\gamma_{\beta}\Theta^{\mu\omega} +
 \Theta^{\alpha\mu}\gamma_{\varphi}\Theta^{\varphi\omega}),
\end{eqnarray}
where the quadratically, $I_1(M_{\rm N})$, and logarithmically,
$I_2(M_{\rm N})$, divergent integrals are determined by [4]
\begin{eqnarray}\label{label2.6}
I_1(M_{\rm N})&=&\int \frac{d^4k}{\pi^2i}\,\frac{1}{M^2_{\rm N} -
k^2}= 2\,\Bigg[\Lambda\,\sqrt{M^2_{\rm N} + \Lambda^2} - M^2_{\rm N}\,{\ell
n}\Bigg(\frac{\Lambda}{M_{\rm N}} +\sqrt{1 +
\frac{\Lambda^2}{M^2_{\rm N}}}\Bigg)\Bigg],\nonumber\\ I_2(M_{\rm
N})&=&\int \frac{d^4k}{\pi^2i}\,\frac{1}{(M^2_{\rm N} - k^2)^2}=
2\,\Bigg[{\ell n}\Bigg(\frac{\Lambda}{M_{\rm N}} +\sqrt{1 +
\frac{\Lambda^2}{M^2_{\rm N}}}\Bigg) - \frac{\Lambda}{\sqrt{M^2_{\rm N} +
\Lambda^2}}\Bigg].
\end{eqnarray}
The cut--off $\Lambda$ restricts from above 3--momenta of fluctuating
nucleon fields. Since we have no closed nucleon loops, the cut--off
$\Lambda$ cannot be determined by the scale of the deuteron size
$r_{\rm D} \sim 1/\Lambda_{\rm D}$ [4]. The natural value of $\Lambda$
is the scale of the Compton wavelength of the nucleon 
$^-\!\!\!\!\lambda_{\rm N}
= 1/M_{\rm N} = 0.21\,{\rm fm}$, i.e. $\Lambda = M_{\rm N}$.

The Lagrangian ${\cal L}^{\rm \Delta\Delta D}_{\rm eff}(x)$ of the
effective ${\rm \Delta \Delta D}$ interaction we obtain in the form
\begin{eqnarray}\label{label2.7}
&&{\cal L}^{\rm \Delta\Delta D}_{\rm eff}(x) = g_{\rm \Delta \Delta D}\,
[\bar{\Delta}^a_{\alpha}(x)\,\Theta^{\alpha\beta}\gamma^{\mu}{\Theta_{\beta}}^{\omega}\tau_2
\Delta^a_{\omega}(x)^c]\,D_{\mu}(x)\nonumber\\
&& + \bar{g}_{\rm \Delta \Delta D}\,[\bar{\Delta}^a_{\alpha}(x)\,(\Theta^{\alpha\beta}\gamma_{\beta}\Theta^{\mu\omega} +
 \Theta^{\alpha\mu}\gamma_{\varphi}\Theta^{\varphi\omega})\,\tau_2
\Delta^a_{\omega}(x)^c]\,D_{\mu}(x)=\nonumber\\
&&=-\,i\,g_{\rm \Delta \Delta
D}[\bar{\Delta}^-_{\alpha}(x)\,\Theta^{\alpha\beta}\gamma^{\mu}{\Theta_{\beta}}^{\omega}
\Delta^{++}_{\omega}(x)^c -
\bar{\Delta}^{++}_{\alpha}(x)\,\Theta^{\alpha\beta}\gamma^{\mu}{\Theta_{\beta}}^{\omega}
\Delta^{-}_{\omega}(x)^c\nonumber\\ && +
\bar{\Delta}^+_{\alpha}(x)\Theta^{\alpha\beta}\gamma^{\mu}{\Theta_{\beta}}^{\omega}
\Delta^0_{\omega}(x)^c -
\bar{\Delta}^0_{\alpha}(x)\Theta^{\alpha\beta}\gamma^{\mu}{\Theta_{\beta}}^{\omega}
\Delta^+_{\omega}(x)^c]\,D_{\mu}(x)\nonumber\\
&&-\,i\,\bar{g}_{\rm \Delta \Delta
D}[\bar{\Delta}^-_{\alpha}(x)\,(\Theta^{\alpha\beta}\gamma_{\beta}\Theta^{\mu\omega} +
 \Theta^{\alpha\mu}\gamma_{\varphi}\Theta^{\varphi\omega})\,
\Delta^{++}_{\omega}(x)^c\nonumber\\
&&-\bar{\Delta}^{++}_{\alpha}(x)\,(\Theta^{\alpha\beta}\gamma_{\beta}\Theta^{\mu\omega} +
 \Theta^{\alpha\mu}\gamma_{\varphi}\Theta^{\varphi\omega})\,
\Delta^{-}_{\omega}(x)^c\nonumber\\
&&+ \bar{\Delta}^{+}_{\alpha}(x)\,(\Theta^{\alpha\beta}\gamma_{\beta}\Theta^{\mu\omega} +
 \Theta^{\alpha\mu}\gamma_{\varphi}\Theta^{\varphi\omega})\,
\Delta^{0}_{\omega}(x)^c\nonumber\\
&&- \bar{\Delta}^{0}_{\alpha}(x)\,(\Theta^{\alpha\beta}\gamma_{\beta}\Theta^{\mu\omega} +
 \Theta^{\alpha\mu}\gamma_{\varphi}\Theta^{\varphi\omega})\,
\Delta^{+}_{\omega}(x)^c],
\end{eqnarray}
where the effective coupling constants $g_{\rm \Delta \Delta D}$ and
$\bar{g}_{\rm \Delta \Delta D}$ read
\begin{eqnarray}\label{label2.8}
g_{\rm \Delta \Delta D} &=& g_{\rm V}\,\frac{7g^2_{\rm \pi
N\Delta}}{384\pi^2}\,\Bigg[\frac{\Lambda}{\sqrt{M^2_{\rm N} +
\Lambda^2}}\,\Bigg(1 + \frac{2}{7}\,\frac{\Lambda^2}{M^2_{\rm
N}}\Bigg) - {\ell n}\Bigg(\frac{\Lambda}{M_{\rm N}} +\sqrt{1 +
\frac{\Lambda^2}{M^2_{\rm N}}}\Bigg)\Bigg],\nonumber\\ \bar{g}_{\rm
\Delta \Delta D} &=&-\, g_{\rm V}\,\frac{g^2_{\rm \pi
N\Delta}}{192\pi^2}\,\Bigg[\frac{\Lambda}{\sqrt{M^2_{\rm N} +
\Lambda^2}}\,\Bigg(1 + \frac{1}{2}\,\frac{\Lambda^2}{M^2_{\rm
N}}\Bigg) - {\ell n}\Bigg(\frac{\Lambda}{M_{\rm N}} +\sqrt{1 +
\frac{\Lambda^2}{M^2_{\rm N}}}\Bigg)\Bigg].
\end{eqnarray}
On--mass shell of the $\Delta(1232)$ resonance, i.e. in the case of
the PMA [1,12], the contribution of the parameter $Z$ vanishes and the
effective ${\rm \Delta \Delta D}$ interaction acquires the form
\begin{eqnarray}\label{label2.9}
&&{\cal L}^{\rm \Delta\Delta D}_{\rm eff}(x) = 
g_{\rm \Delta \Delta
D}\,g^{\alpha\beta} [\bar{\Delta}^a_{\alpha}(x)\,\gamma^{\mu}\tau_2
\Delta^a_{\beta}(x)^c]\,D_{\mu}(x)=\nonumber\\ 
&&=-\,i\,g_{\rm \Delta
\Delta D}\,g^{\alpha\beta}\,[\bar{\Delta}^-_{\alpha}(x)\gamma^{\mu}
\Delta^{++}_{\beta}(x)^c -
\bar{\Delta}^{++}_{\alpha}(x)\gamma^{\mu}
\Delta^{-}_{\beta}(x)^c\nonumber\\ && +
\bar{\Delta}^+_{\alpha}(x)\gamma^{\mu}
\Delta^0_{\beta}(x)^c -
\bar{\Delta}^0_{\alpha}(x)\gamma^{\mu}
\Delta^+_{\beta}(x)^c]\,D_{\mu}(x).
\end{eqnarray}
The total probability $P(\Delta\Delta)$ to find the $\Delta\Delta$
component inside the deuteron we determine as follows
\begin{eqnarray}\label{label2.10}
P(\Delta\Delta) = \frac{d\Gamma({\rm D \to
\Delta\Delta})}{d\Gamma({\rm D \to n p})},
\end{eqnarray}
where $d\Gamma({\rm D \to \Delta\Delta})$ and $d\Gamma({\rm D \to n
p})$ are the differential rates of the transitions D $\to$ $\Delta$ +
$\Delta$ and D $\to$ n + p, respectively, defined by
\begin{eqnarray}\label{label2.11}
&&d\Gamma({\rm D}(P) \to \Delta(p_1)\Delta(p_2))= 8\,g^2_{\rm
\Delta \Delta
D}\,\frac{d\Phi_{\Delta\Delta}(p_1,p_2)}{6\,\sqrt{s}}\,\Bigg(-\,g_{\mu\nu}
+ \frac{P_{\mu}P_{\nu}}{s}\Bigg)\nonumber\\ &&\times\,{\rm
tr}\Bigg\{(M_{\Delta} + \hat{p}_1)\,\Bigg(-\,g_{\alpha\beta} +
\frac{1}{3}\,\gamma_{\alpha}\gamma_{\beta} +
\frac{1}{3}\,\frac{\gamma_{\alpha}p_{1\beta} -
\gamma_{\beta}p_{1\alpha}}{M_{\Delta}} +
\frac{2}{3}\,\frac{p_{1\alpha}p_{1\beta}}{M^2_{\Delta}}\Bigg)\,\gamma^{\mu}
\nonumber\\ &&\times\Bigg(-\,g^{\alpha\beta} +
\frac{1}{3}\,\gamma^{\beta}\gamma^{\alpha} +
\frac{1}{3}\,\frac{\gamma^{\beta}p^{\alpha}_2 -
\gamma^{\alpha}p^{\beta}_2}{M_{\Delta}} +
\frac{2}{3}\,\frac{p^{\beta}_2p^{\alpha}_2}{M^2_{\Delta}}\Bigg)\,(-\,M_{\Delta}
+ \hat{p}_2)\gamma^{\nu}\Bigg\},\nonumber\\ &&
d\Gamma({\rm D}(P) \to
{\rm n}(p_1) {\rm p}(p_2)) = 4\,g^2_{\rm V}\,\frac{d\Phi_{\rm n
p}(p_1,p_2)}{6\,\sqrt{s}}\,\Bigg(-\,g_{\mu\nu} +
\frac{P_{\mu}P_{\nu}}{s}\Bigg)\nonumber\\ &&\times\,{\rm tr}\{(M_{\rm
N} + \hat{p}_1)\gamma^{\mu}(-\,M_{\rm N} + \hat{p}_2)\gamma^{\nu}\}.
\end{eqnarray}
We have denoted as $P = p_1 + p_2$ and $P^2 = s$ the 4--momentum and
the invariant squared mass of the deuteron, respectively. Then,
$d\Phi_{\Delta\Delta}(p_1,p_2)$ and $d\Phi_{\rm n p}(p_1,p_2)$ are the
phase volumes of the $\Delta\Delta$ and np states. The two--particle
phase volume is equal to
\begin{eqnarray}\label{label2.12}
d\Phi(p_1,p_2) = (2\pi)^4(P - p_1 - p_2)\,\frac{d^3 p_1}{(2\pi)^32
E_1}\,\frac{d^3 p_2}{(2\pi)^3 2 E_2}.
\end{eqnarray}
At leading order in the large $N_C$ expansion, when we can neglect the
mass difference between the $\Delta(1232)$ resonance and the nucleon,
the phase volumes $d\Phi_{\Delta\Delta}(p_1,p_2)$ and
$d\Phi_{\rm n p}(p_1,p_2)$ are equal
\begin{eqnarray}\label{label2.13}
d\Phi_{\Delta\Delta}(p_1,p_2) = d\Phi_{\rm n p}(p_1,p_2) =
d\Phi(p_1,p_2).
\end{eqnarray}
The differential rates $d\Gamma({\rm D}(P) \to
\Delta(p_1)\Delta(p_2))$ and $d\Gamma({\rm D}(P) \to {\rm
n}(p_1) {\rm p}(p_2))$ calculated at leading order in the large $N_C$
expansion are given by
\begin{eqnarray}\label{label2.14}
d\Gamma({\rm D}(P) \to \Delta(p_1)\Delta(p_2)) &=&
\frac{10}{9}\times 8 \times g^2_{\rm \Delta \Delta
D} \times \sqrt{s}\,d\Phi(p_1,p_2),\nonumber\\ d\Gamma({\rm D}(P) \to {\rm
n}(p_1) {\rm p}(p_2)) &=& 4\times g^2_{\rm V} \times
\sqrt{s}\,d\Phi(p_1,p_2).
\end{eqnarray}
Hence, the probability $P(\Delta\Delta)$ to find the $\Delta\Delta$
component inside the deuteron amounts to
\begin{eqnarray}\label{label2.15}
P(\Delta\Delta) = \frac{10}{9}\times \frac{2 g^2_{\rm \Delta \Delta
D}}{g^2_{\rm V}} = 0.3\,\%,
\end{eqnarray}
where the numerical value is obtained at $\Lambda = M_{\rm N}$.

Our theoretical prediction agrees good with recent experimental
estimate of the upper limit $P(\Delta\Delta) < 0.4\%$ at 90$\%$ of CL
[14] quoted by Dymarz and Khanna [1].

\section{Conclusion}

The theoretical estimate of the contribution of the $\Delta\Delta$
component to the deuteron obtained in the NNJL model agrees good with
the experimental upper limit. Indeed, for the $\Delta(1232)$ resonance
on--mass shell [1,12] we predict $P(\Delta\Delta) = 0.3\,\%$ whereas
experimentally $P(\Delta\Delta)$ is restricted by $P(\Delta\Delta) <
0.4\%$ at 90$\%$ of CL [14].

Off--mass shell of the $\Delta(1232)$ resonance, where the parameter
$Z$ should contribute, our prediction for $P(\Delta\Delta)$ can be
changed, of course. Moreover, due to $Z$ dependence the contributions
of the $\Delta\Delta$ component to amplitudes of different low--energy
nuclear reactions and physical quantities can differ each
other. However, we would like to emphasize that in the NNJL model by
using the effective ${\rm \Delta \Delta D}$ interaction determined by
Eq.(\ref{label2.7}) one can calculate the contribution of the
$\Delta\Delta$ component of the deuteron to the amplitude of any
low--energy nuclear reaction with the deuteron in the initial or final
state.

In our approach we do not distinguish contributions of the
$\Delta\Delta$--pair with a definite orbital momentum ${^3}{\rm
S^{\Delta\Delta}_1}$, ${^3}{\rm D^{\Delta\Delta}_1}$ and so on to the
effective ${\rm \Delta \Delta D}$ interaction Eq.(\ref{label2.7}).
The obtained value of the probability $P(\Delta\Delta)$ should be
considered as a sum of all possible states with a certain orbital
momentum.

Our prediction $P(\Delta\Delta) = 0.3\,\%$ agrees reasonably well
with the result obtained by Dymarz and Khanna in the PMA [1]:
$P(\Delta\Delta) \simeq 0.4 \div 0.5\,\%$.  Unlike our approach Dymarz
and Khanna have given a percentage of the probabilities of different
states ${^3}{\rm S^{\Delta\Delta}_1}$, ${^3}{\rm D^{\Delta\Delta}_1}$
and so to the wave function of the deuteron. In our approach the
deuteron couples to itself and other particles through the one--baryon
loop exchanges. The effective Lagrangian ${\cal L}^{\rm \Delta\Delta
D}_{\rm eff}(x)$ of the ${\rm \Delta\Delta D}$ interaction given by
Eq.(\ref{label2.7}) defines completely the contribution of the
$\Delta\Delta$ intermediate states to baryon--loop exchanges. The
decomposition of the effective ${\rm \Delta\Delta D}$ interaction in
terms of the $\Delta\Delta$ states with a certain orbital momentum
should violate Lorentz invariance for the evaluation of the
contribution of every state to whether the amplitude of a low--energy
nuclear reaction or a low--energy physical quantity. In the NNJL model
this can lead to incorrect results. The relativistically covariant
procedure of the decomposition of the interactions like the ${\rm
\Delta\Delta D}$ one in terms of the states with a certain orbital
momenta is now in progress in the NNJL model. However, a smallness of
the contribution of the $\Delta\Delta$ component to the deuteron
obtained in the NNJL model makes such a decomposition applied to the
${\rm \Delta\Delta D}$ interaction meaningless to some extent due to
impossibility to measure the terms separately.

\section{Acknowledgement}

We are grateful to Prof. W. Plessas for discussions which stimulated
this investigation.

\end{document}